\documentclass[11pt, a4paper]{article}
\usepackage{amsmath}
\newtheorem{theorem}{Theorem}

\begin{document}

\title{On a class of three-dimensional integrable
Lagrangians}
\author{E.V.
Ferapontov, K.R. Khusnutdinova\thanks{On leave from: Institute of Mechanics,
Ufa Branch of the Russian Academy of Sciences, Karl Marx Str. 6, Ufa, 450000,
Russia.}
\  and S.P. Tsarev\thanks{SPT acknowledges partial
financial support from the RFBR grant 04-01-00130 and KSPU grant 4-04-1/FK.}
}
   \date{}
\vspace{-20mm}
   \maketitle
\vspace{-7mm}
\begin{center}
Department of Mathematical Sciences \\ Loughborough
University \\
Loughborough, Leicestershire LE11 3TU, UK \\[1ex] 
and \\[1ex]
Krasnoyarsk State Pedagogical University\\
89 Lebedevoi, 660049 Krasnoyrsk, Russia\\[1ex]
e-mails: \\
\texttt{E.V.Ferapontov@lboro.ac.uk}\\
\texttt{K.Khusnutdinova@lboro.ac.uk}\\
\texttt{tsarev@newmail.ru}\\
\end{center}

\medskip

\begin{abstract}
We characterize non-degenerate Lagrangians of the form
 $$
 \int f(u_x, u_y, u_t) \, dx\, dy\, dt
 $$
such that the corresponding Euler-Lagrange equations
 $
 (f_{u_x})_x+ (f_{u_y})_y+ (f_{u_t})_t=0
 $
are integrable by the method of hydrodynamic reductions.
The integrability conditions
constitute an over-determined system of fourth order PDEs for the Lagrangian
density $f$, which is in involution and possess interesting differential-geometric properties. The moduli space of  integrable Lagrangians,
factorized by the action of a natural equivalence group, is three-dimensional.
Familiar examples include the dispersionless Kadomtsev-Petviashvili (dKP) and the Boyer-Finley
Lagrangians, $f=u_x^3/3+u_y^2-u_xu_t$ and $f=u_x^2+u_y^2-2e^{u_t}$, respectively.  A
complete description of integrable cubic and quartic Lagrangians is obtained. Up to  the equivalence transformations, the list of integrable cubic Lagrangians reduces to three examples,
$$
f=u_xu_yu_t, ~~~  f=u_x^2u_y+u_yu_t ~~~ {\rm and} ~~~ f=u_x^3/3+u_y^2-u_xu_t  \ ({\rm dKP}).
$$
There exists a unique integrable quartic Lagrangian,
$$
f=u_x^4+2u_x^2u_t-u_xu_y-u_t^2.
$$
We conjecture that these examples exhaust the list of integrable polynomial Lagrangians which are essentially three-dimensional (it was  verified  that there exist no polynomial integrable Lagrangians of degree five).

We prove that the Euler-Lagrange equations are integrable by the method of
hydrodynamic reductions if and only if they possess
a scalar pseudopotential playing the role of  a dispersionless `Lax pair'.

\medskip


\medskip

\end{abstract}

\newpage

 \section{Introduction}

The method of hydrodynamic reductions, originally proposed for the dKP equation
in \cite{Gibb94, GibTsa96,  GibTsa99}, applies to a broad class of multi-dimensional
dispersionless PDEs
$F(u, u_s, \ldots )=0$
where $u$ is a
(vector-) function  of $d$ independent variables $s=(x, y, t, \ldots )$.
The main idea is to look for  solutions in the form
${ u}={ u}(R^1, \ldots ,
R^n)$  where the Riemann invariants
$R^1(x,y,t,\ldots)$, \ldots, $R^n(x,y,t,\ldots)$ are arbitrary solutions of
$(d-1)$
commuting
diagonal systems
$$
R^i_y=\mu^i(R)\ R^i_x, \ \ R^i_t=\lambda^i(R)\ R^i_x,
$$
etc. We recall, see \cite{Tsarev}, that the
commutativity conditions
imply the following restrictions on the
characteristic speeds:
\begin{equation}
\frac{\partial_j\mu^i}{\mu^j-\mu^i}=\frac{\partial_j\lambda
^i}{\lambda^j-\lambda^i}, ~~~
i\ne j, ~~ \partial_i=\partial/\partial_{R^i}.
\label{comm}
\end{equation}
Thus, the original  multi-dimensional equation  is
decoupled into a collection
of commuting $n$-component  $(1+1)$-dimensional systems in Riemann invariants
which  can then be solved by the generalized hodograph method \cite{Tsarev}.
Solutions arising within this approach, known as nonlinear interactions of  $n$
planar
simple waves, were extensively investigated in gas dynamics (simple
waves, double waves, etc, \cite{Sidorov}) and the theory of dispersionless
$(2+1)$-dimensional  hierarchies of the dKP type
\cite{Gibb94, GibTsa96,
GibTsa99, GuMaAl, Ma, Man, M,  Manas, Fer,  Shabat, Kr3, Pavlov}.

It was suggested in \cite{Fer4, Fer6} to call a
multi-dimensional equation {\it
integrable} if it possesses
`sufficiently many' $n$-component hydrodynamic
reductions  (parametrized by $(d-2)n$
arbitrary functions of a single variable). It turned out that this requirement
is very strong and provides the effective classification criterion. Partial
classification results for $(2+1)$-dimensional integrable systems of
hydrodynamic
type,
$$
{\bf u}_t+A({\bf u}){\bf u}_x+B({\bf u}){\bf u}_y=0,
$$
(here ${\bf u}$ is $m$-component vector, $A$ and $B$ are $m\times m$ matrices) were
obtained in \cite{Fer4, Fer5}. It was observed that the class of PDEs amenable
to the method of hydrodynamic reductions extends beyond the class of
hydrodynamic type systems. For instance, the method applies to scalar second
order PDEs
$$
F(u_{xx}, u_{xy}, u_{xt}, u_{yy}, u_{yt}, u_{tt})=0,
$$
see e.g. \cite{Pavlov1}, \cite{FK-appear} for the classification of integrable
equations of the
form $u_{tt}=f(u_{xx}, u_{xt}, u_{xy})$. Hydrodynamic reductions of
multi-dimensional dispersionless integrable systems in dimensions
greater then three were discussed in \cite{Fer2, Fer6}.

In this paper we apply the method of hydrodynamic reductions to
three-dimensional equations of the form
\begin{equation}
 (f_{u_x})_x+ (f_{u_y})_y+ (f_{u_t})_t=0,
\label{zero}
\end{equation}
 which are the Euler-Lagrange equations corresponding to first order Lagrangian densities $ f(u_x, u_y, u_t)$. We begin with two illustrative examples.

\medskip

{\bf Example 1.} Consider the linear wave equation $u_{tt}=u_{xx}+u_{yy}$ corresponding to the quadratic Lagrangian density $f=u_x^2+u_y^2-u_t^2$. Setting $a=u_x, \ b=u_y, \ c=u_t$ we can  rewrite it in the first order form
\begin{equation}
a_y=b_x, ~~~ a_t=c_x, ~~~ b_t=c_y, ~~~ a_x+b_y-c_t=0.
\label{wave}
\end{equation}
Let us seek solutions  in the form
$a=a(R^1, \ldots ,
R^n), \ b=b(R^1, \ldots , R^n), \ c=c(R^1, \ldots , R^n)$ where the  Riemann
invariants
$R^1(x,y,t)$, \ldots , $R^n(x,y,t)$ are {\it arbitrary} solutions of
 a pair of commuting hydrodynamic type systems
$$
 R^i_y=\mu^i(R)\ R^i_t,  ~~~~ R^i_x=\lambda^i(R)\ R^i_t;
$$
(for the sake of symmetry we have chosen $t$ as a distinguished variable). Substituting this ansatz into (\ref{wave}) one obtains the equations
$\partial_ib=\mu^i\partial_ic, \  \partial_ia=\lambda^i\partial_ic$
 along with the dispersion relation
$$
(\lambda^i)^2+(\mu^i)^2=1.
$$
Setting $\lambda^i=\cos \varphi^i, \ \mu^i=\sin \varphi^i$ and taking into account the commutativity conditions (\ref{comm}) one obtains $\partial_j\varphi^i=0$ for any $j\ne i.$ Thus, $\varphi^i=\varphi^i(R^i)$ so that both systems of hydrodynamic type assume  decoupled forms
$$
 R^i_y=\sin \varphi^i(R^i)\ R^i_t,  ~~~~ R^i_x=\cos \varphi^i(R^i)\ R^i_t,
 $$
 with the general solution $R^i(x, y, t)$ given by the implicit  relation
 \begin{equation}
 g^i(R^i)=t+y \sin \varphi^i(R^i)+x \cos \varphi^i(R^i);
 \label{imp}
 \end{equation}
 here $g^i(R^i)$ are yet another $n$ arbitrary functions. Finally,  the consistency conditions for the equations $\partial_ib=\mu^i\partial_ic, \  \partial_ia=\lambda^i\partial_ic$ imply $\partial_i\partial_j c=0$ so that one can set $c=R^1+...+R^n$ (notice that one has a reparametrization freedom $R^i\to f^i(R^i)$).
 Thus, we arrive at solutions of the wave equation given by the formulae
 $$
 \begin{array}{c}
 u_t=c=R^1+... +R^n, \\
 \ \\
 u_y=b=\int \sin \varphi^1(R^1) dR^1+ ... +\int \sin \varphi^n(R^n) dR^n, \\
 \ \\
 u_x=a=\int \cos \varphi^1(R^1) dR^1+ ... +\int \cos \varphi^n(R^n) dR^n,
 \end{array}
 $$
 where  $R^i(x,  y,  t)$ is defined by the implicit relation (\ref{imp}).  These solutions depend on $2n$ arbitrary functions of a single argument and can be viewed as  linear superpositions of $n$ elementary solutions of the form
 $$
 \begin{array}{c}
  u_t=R, ~~~
 u_y=\int \sin \varphi (R) dR, ~~~
 u_x=\int \cos \varphi(R) dR
 \end{array}
$$
where  $R(x, y, t)$ is defined by the implicit relation
$ g(R)=t+y \sin \varphi (R)+x \cos \varphi (R)$.
This relation implies that
 the level sets $R=const$ are null-planes so that the gradient of $u$ is constant along a one-parameter family of null-planes. Solutions of this type are known as planar simple waves.

 \medskip

 {\bf Example 2.} Let us apply the same approach to  the {\it nonlinear} wave  equation $e^{u_t}u_{tt}=u_{xx}+u_{yy}$
 (also known as the Boyer-Finley equation) which corresponds to the  Lagrangian density $f=u_x^2+u_y^2-2e^{u_t}$. Setting $a=u_x, \ b=u_y, \ c=u_t$ we can rewrite it as follows:
\begin{equation}
a_y=b_x, ~~~ a_t=c_x, ~~~ b_t=c_y, ~~~ a_x+b_y-e^c c_t=0.
\label{nwave}
\end{equation}
We again seek solutions  in the form
$a=a(R^1, \ldots ,
R^n), \ b=b(R^1, \ldots , R^n), \ c=c(R^1, \ldots , R^n)$ where the  Riemann
invariants solve hydrodynamic type systems
$$
 R^i_y=\mu^i(R)\ R^i_t,  ~~~~ R^i_x=\lambda^i(R)\ R^i_t.
$$
The substitution  into (\ref{nwave}) implies
\begin{equation}
\partial_ib=\mu^i\partial_ic, \ \ \  \partial_ia=\lambda^i\partial_ic
\label{c}
\end{equation}
 along with the dispersion relation
$$
(\lambda^i)^2+(\mu^i)^2=e^c.
$$
Setting $\lambda^i=e^{c/2}\cos \varphi^i, \ \mu^i=e^{c/2}\sin \varphi^i$ and taking into account the commutativity conditions (\ref{comm}) one obtains the expression for $\partial_j \varphi^i$:
\begin{equation}
\partial_j \varphi^i=-\frac{1}{2}\cot \frac{\varphi^i-\varphi^j}{2} \ \partial_j c,  ~~~ i\ne j.
\label{phi}
\end{equation}
The consistency conditions for the equations (\ref{c})  imply
$$
\partial_i\partial_jc=\frac{\partial_j\lambda
^i}{\lambda^j-\lambda^i}\partial_i
c+\frac{\partial_i\lambda
^j}{\lambda^i-\lambda^j}\partial_j c, ~~~ i\ne j.
$$
Substituting here $\lambda^i=e^{c/2}\cos \varphi^i$ and taking into account (\ref{phi}) we obtain the following system which governs hydrodynamic reductions of the nonlinear wave equation:
\begin{equation}
\partial_j \varphi^i=-\frac{1}{2}\cot \frac{\varphi^i-\varphi^j}{2} \ \partial_j c, ~~~
\partial_j\partial_i c=\frac{\partial_jc\partial_ic}{2\sin^2\frac{\varphi^i-\varphi^j}{2}};
\label{bf}
\end{equation}
notice that this system is a nonlinear analog of the system $\partial_j \varphi^i=0, \
\partial_j\partial_i c=0$ derived in  Example 1 for the linear wave equation.
One can show that the system (\ref{bf}) is in involution (that is, the compatibility conditions $\partial_k(\partial_j\varphi^i)=\partial_j(\partial_k\varphi^i)$ and
$\partial_k(\partial_j\partial_ic)=\partial_j(\partial_k\partial_ic)$ are satisfied identically)
and its general solution depends, modulo reparametrizations $R^i\to f^i(R^i)$, on $n$ arbitrary functions of a single argument. Once a solution to the system (\ref{bf}) is found, one can reconstruct $b(R)$ and $a(R)$ from the equations (\ref{c}) which are consistent by construction. After that one has to solve  hydrodynamic type systems (where the characteristic speeds $\mu^i(R)$ and $\lambda^i(R)$ are known) which can be done by the generalized hodograph method \cite{Tsarev}. This gives some more $n$ arbitrary functions. Thus, solutions arising from $n$-component reductions of the nonlinear wave equation depend on $2n$ arbitrary functions of a single argument.
For one-component reductions, equations (\ref{bf}) become vacuous and, setting $c=R$, one has
 $$
 \begin{array}{c}
  u_t=c=R, ~~~
 u_y=b=\int e^{R/2} \sin \varphi (R) dR, ~~~
 u_x=a=\int e^{R/2} \cos \varphi(R) dR
 \end{array}
$$
where the function $R(x, y, t)$ is defined by the implicit relation
$ g(R)=t+y e^{R/2} \sin \varphi (R)+x e^{R/2} \cos \varphi (R)$.
This relation implies that the level sets $R=const$ are characteristic planes. Solutions of this type are known as planar simple waves.
Thus, solutions governed by the system (\ref{bf})  can  be interpreted as {\it nonlinear interactions} of planar simple waves.
We refer to \cite{Fer, Fer2, M, Man} for  further discussion and explicit examples.

\bigskip

The main object of our study are nonlinear Lagrangian PDEs (of the type discussed in Example 2) which are integrable in the above sense, that is, possess infinitely many hydrodynamic reductions. This requirement imposes strong restrictions on the Lagrangian density $f$.
Our first main result is the system of partial differential equations for the
density $f(a, b, c)$
providing the necessary and sufficient conditions for the integrability (we use
the notation $a=u_x, \ b=u_y, \ c=u_t$). These conditions  can be
represented in a remarkable
compact form:

\begin{theorem}  For  non-degenerate Lagrangians, the Euler-Lagrange equations
(\ref{zero})
 are integrable by the method of hydrodynamic reductions if
and only if the density $f$ satisfies the identity
\begin{equation}
d^4f=d^3f\frac{dH}{H}+\frac{3}{H}{ det} (dM).
\label{fourth}
\end{equation}
\end{theorem}
Here $d^3f$ and $d^4f$ are the symmetric differentials of $f$ which appear in
the standard Taylor expansion $f(s+ds)-f(s)=df+d^2f/2!+d^3f/3!+\ldots $ for a
function $f(s)$ of three variables $s=(a, b, c)$. The Hessian $H$ and the $4 \times
4$ matrix $M$ are defined as follows:
\begin{equation}
H=det
\left(\begin{array}{ccc}
f_{aa} & f_{ab} & f_{ac} \\
f_{ab} & f_{bb} & f_{bc} \\
f_{ac} & f_{bc} & f_{cc}
\end{array}
\right), ~~~
M=\left(\begin{array}{cccc}
0 & f_a & f_b & f_c \\
f_a & f_{aa} & f_{ab} & f_{ac} \\
f_b & f_{ab} & f_{bb} & f_{bc} \\
f_c & f_{ac} & f_{bc} & f_{cc}
\end{array}
\right).
\label{Hessian}
\end{equation}
The differential $dM=M_ada+M_bdb+M_cdc$ is a matrix-valued  linear form
$$
\left(\begin{array}{cccc}
0 & f_{aa} & f_{ab} & f_{ac} \\
f_{aa} & f_{aaa} & f_{aab} & f_{aac} \\
f_{ab} & f_{aab} & f_{abb} & f_{abc} \\
f_{ac} & f_{aac} & f_{abc} & f_{acc}
\end{array}
\right)da+
\left(\begin{array}{cccc}
0 & f_{ab} & f_{bb} & f_{bc} \\
f_{ab} & f_{aab} & f_{abb} & f_{abc} \\
f_{bb} & f_{abb} & f_{bbb} & f_{bbc} \\
f_{bc} & f_{abc} & f_{bbc} & f_{bcc}
\end{array}
\right)db+
\left(\begin{array}{cccc}
0 & f_{ac} & f_{bc} & f_{cc} \\
f_{ac} & f_{aac} & f_{abc} & f_{acc} \\
f_{bc} & f_{abc} & f_{bbc} & f_{bcc} \\
f_{cc} & f_{acc} & f_{bcc} & f_{ccc}
\end{array}
\right)dc.
$$
A Lagrangian is said to be non-degenerate iff $H \ne 0$.

Both sides of the equation (\ref{fourth}) are homogeneous symmetric quartic
forms in commuting formal variables $da, db,
dc$. Equating similar terms we obtain expressions for {\it all} fourth order partial
derivatives of $f$ in terms of its second and third order derivatives (15 equations altogether). The
resulting over-determined system for $f$ is in involution, and its solution
space is $20$-dimensional (indeed, the values of the derivatives of $f$ up to
order 3 at a given point
$(a_0, b_0, c_0)$  amount to  $20$ arbitrary constants). Factorised  by the action of
a natural
equivalence group of dimension 17 (generated by arbitrary affine transformations of the independent variables $a, b, c$ plus transformations $f\to \mu f+\alpha a+\beta b+\gamma c+\delta$, see the end of Sect. 2 for the discussion of the origin of these symmetries) this provides a
three-dimensional moduli space of integrable Lagrangians. Details of the
derivation of the integrability conditions (\ref{fourth}) are given in Sect. 2.
Notice that  these conditions arise in a somewhat different form, namely, as
explicit formulae for the fourth order derivatives of $f$ (which are very
complicated). It is a truly remarkable fact that they compactify into a single
expression (\ref{fourth}). Differential-geometric aspects and an invariant formulation of the integrability conditions (\ref{fourth}) are discussed in Sect. 5.

{\bf Example 3}. For the dKP Lagrangian ($f=a^3/3+b^2-ac$) one has $d^4f=0, \
H=-2, \ {\rm det} (dM)=0$.
Similarly, for the Boyer-Finley  Lagrangian ($f=a^2+b^2-2e^c$) one has
$d^4f=-2e^c(dc)^4, \ d^3f=-2e^c(dc)^3, \ H=-2e^c, \ {\rm det} (dM)=0$. In both
cases the identity (\ref{fourth}) is obviously satisfied.

{\bf Remark.} Notice that in two dimensions any Euler-Lagrange equation of
the form $(f_{u_x})_x+(f_{u_y})_y=0$ is automatically integrable. Indeed, in the
new variables $a=u_x, \ b=u_y$
it takes the form of a two-component quasilinear system $a_y=b_x, \
(f_a)_x+(f_b)_y=0$ which linearises under the hodograph transformation
interchanging dependent and independent variables. This trick, however, does not work in more than two dimensions.

\medskip

Sect. 3 is devoted to polynomial Lagrangians. Our first result is a simple Lemma stating that, for non-degenerate homogeneous solutions of the equation (\ref{fourth}), the degree of homogeneity can take only one of the three values $0,  2$ or $3$ (recall that Lagrangians of homogeneity one are automatically degenerate). This observation is particularly useful for the classification of polynomial Lagrangians implying that the `leading'  homogeneous part of a polynomial solution $f$ must be either  of degree $3$ or  degenerate (polynomials of degree two give rise to linear Euler-Lagrange equations).
In Sect. 3.1 we obtain a complete list of integrable cubic Lagrangians with the densities
$$
f=C(u_x, u_y, u_t)+\alpha u_x^2+\beta u_y^2+\gamma u_t^2+ \mu u_xu_y +\nu u_xu_t
+\eta u_yu_t;
$$
here $C$ is a homogeneous cubic form. The substitution into the integrability
conditions (\ref{fourth}) implies that $C$ must be totally reducible, leading to three
essentially different possibilities:

\noindent (1) $C$ consists of three lines in a general position. Further
analysis allows one to  eliminate  quadratic terms, leading to a  unique
Lagrangian density
$$
f=u_xu_yu_t
$$
with the corresponding Euler-Lagrange equation
$u_tu_{xy}+u_yu_{xt}+u_xu_{yt}=0.$ Although this example looks deceptively simple, the corresponding dispersionless Lax pair is quite non-trivial, see formula (\ref{z1}). To the best of our knowledge this  Lagrangian has not been recorded before.

\noindent (2) $C$ contains a double line. Up to  equivalence transformations
this case reduces to the density
$$
f=u_x^2u_y+u_yu_t
$$
which generates a particular flow of the so-called $r$-Dym hierarchy.

\noindent (3) $C$ is a triple line. This case reduces to the dKP Lagrangian
density
$$
f=u_x^3/3+u_y^2-u_xu_t.
$$

\medskip

Quartic Lagrangians are classified in Sect. 3.2. Here the leading part of the  density $f$ must necessarily be degenerate, leading  to a unique integrable example
$$
f=u_x^4+2u_x^2u_t-u_xu_y-u_t^2
$$
which corresponds to a particular flow of the so-called $r$-dKP hierarchy.

We have verified that there exist no polynomial solutions to the system (\ref{fourth}) of  degree five. It is tempting to conjecture that the four examples listed above exhaust the list of polynomial integrable Lagrangians.

\medskip

Homogeneous Lagrangians can be obtained by setting $f(a, b, c)=a^kg(\xi, \eta), \ \xi={b}/{a}, \ \eta={c}/{a}$ (recall that the degree of homogeneity $k$ can take the values $0, 2, 3$ only). Substituting this ansatz into the integrability conditions (\ref{fourth}) and analysing the resulting equations for $g(\xi, \eta)$ one can show   that the case $k=3$ leads to the cubic Lagrangian $f=u_xu_yu_t$ and, thus, gives no new examples. The case $k=2$ reduces to functions $g$ which are quadratic in $\xi, \eta$ and, hence,   generate quadratic Lagrangians with linear Euler-Lagrange equations.
The last  case $k=0$ turned out to be quite nontrivial. The substitution of $f(a, b, c)=g(\xi, \eta)$ into (\ref{fourth}) results in a system of five equations expressing all fourth order partial derivatives of $g$ in terms of lower order derivatives. In symbolic form, this system can be represented as follows:
\begin{equation}
d^4g=d^3g\frac{dh}{h}+6\frac{dg}{h}{ det} (dm)+3\frac{(dg)^2}{h}det (dn).
\label{fourth1}
\end{equation}
Here, as in (\ref{fourth}),  $d^sg$  are  symmetric differentials of $g$ (notice that $g$ is now a function of  two variables), the matrices $m$ and $n$ are defined as
$$
m=\left(\begin{array}{ccc}
0 & g_{\xi} & g_{\eta}\\
g_{\xi} & g_{\xi \xi} & g_{\xi \eta}  \\
g_{\eta} & g_{\xi \eta} & g_{\eta \eta}
\end{array}
\right), ~~~~
n=\left(\begin{array}{cc}
g_{\xi \xi} & g_{\xi \eta}  \\
g_{\xi \eta} & g_{\eta \eta}
\end{array}
\right),
$$
and
$$
h=-det (m)=g_{\eta}^2g_{\xi \xi}-2g_{\xi}g_{\eta}g_{\xi \eta}+g_{\xi}^2g_{\eta \eta}.
$$
The non-degeneracy of the Lagrangian density $f(a, b, c)=g(\xi, \eta)$   is equivalent to the condition $h\ne 0$. One can show that the over-determined system (\ref{fourth1})  is in involution and its solution
space is $10$-dimensional (indeed, the values of partial derivatives of $g$ up to
order 3 at a fixed point $(\xi_0, \eta_0)$  amount to  $10$ arbitrary constants). Although it is still difficult to integrate this system in general, some particular solutions can readily be constructed. For instance, any quadratic form $g(\xi, \eta)=\alpha \xi^2+\beta \xi \eta +\gamma \eta^2+\mu \xi+\nu \eta$ is a solution of
(\ref{fourth1}), leading to integrable Lagrangians with the densities
$$
f=\alpha \frac{u_y^2}{u_x^2}+\beta \frac{u_tu_y}{u_x^2} +\gamma \frac{u_t^2}{u_x^2}+\mu \frac{u_y}{u_x}+\nu \frac{u_t}{u_x}.
$$
Up to the equivalence transformations  any expression of this type can be reduced to either of the non-equivalent (over reals) canonical forms,
$$
f=\frac{u_y^2+u_t^2}{u_x^2}, ~~~~ f=\frac{u_yu_t}{u_x^2}, ~~~~ f=\frac{u_y^2}{u_x^2}+\frac{u_t}{u_x}.
$$

\medskip

In Sect. 4 we study scalar pseudopotentials for integrable Euler-Lagrange equations (also known as $S$-functions,
dispersionless Lax pairs, etc). Examples
thereof include the dispersionless Lax pair
$$
S_t=\frac{1}{3}S_x^3+u_xS_x+ u_y, ~~~ S_y=\frac{1}{2}S_x^2+ u_x
$$
which generates  the dKP equation $u_{xt}-u_xu_{xx}=u_{yy}.$
Similarly, the Boyer-Finley equation $u_{xy}=(e^{u_t})_t $  possesses the
dispersionless Lax pair
$$
S_t=u_t-\ln S_y, ~~~ S_x=u_x-\frac{e^{u_t}}{  S_y}.
$$
Further examples of integrable multi-dimensional equations
possessing
pseudopotentials of the above type can be found in
\cite{Zakharov, Pavlov1,
Kon}. It was proved in \cite{Fer5} that for two-component $(2+1)$-dimensional
systems of hydrodynamic type the existence of dispersionless Lax pairs is
necessary and sufficient  for the integrability (that is, for the existence of
sufficiently many hydrodynamic reductions).
Dispersionless Lax pairs constitute a basis for the
twistor
and dispersionless $\bar \partial$-approaches to multi-dimensional
dispersionless hierarchies \cite{Mason, Kon, Bogdanov}.

Our second main result is the following

\begin{theorem} The Euler-Lagrange equation
 $(f_{u_x})_x+ (f_{u_y})_y+ (f_{u_t})_t=0$ is integrable by the method of
hydrodynamic reductions if and only if it possesses a dispersionless Lax pair
$$
S_t=F(S_x, \ u_x, \ u_y, \ u_t), ~~~ S_y=G(S_x, \ u_x, \ u_y, \ u_t).
$$
\end{theorem}

\noindent In some cases it seems to be more convenient  to work
with parametric Lax pairs
$$
S_t=F(p, \ u_x, \ u_y, \ u_t), ~~~ S_y=G(p, \ u_x, \ u_y, \ u_t), ~~~ S_x=H(p, \
u_x, \ u_y, \ u_t),
$$
which take the above form if one expresses the parameter $p$ in terms of
$S_x$ from the third equation. For instance, the equation
$u_tu_{xy}+u_yu_{xt}+u_xu_{yt}=0$ corresponding to the Lagrangian density $f=u_xu_yu_t$ possesses the parametric Lax pair
\begin{equation}
\frac{S_x}{u_x}=\zeta(p), ~~~
\frac{S_y}{u_y}=\zeta(p)+\frac{\wp'(p)+\lambda}{2\wp(p)}, ~~~
 \frac{S_t}{u_t}=\zeta(p)+\frac{\wp'(p)-\lambda}{2\wp(p)};
 \label{z1}
 \end{equation}
here $(\wp')^2=4\wp^3+\lambda^2$ and $\zeta'=-\wp$  (Weierstrass $\wp$ and
$\zeta$ functions, see Sect. 3).
\bigskip

\medskip

Differential-geometric aspects of the integrability conditions (\ref{fourth}) are investigated in Sect. 5.
The main object associated with the Lagrangian density $f(a, b, c)$ is the Hessian metric $d^2f$.
We show that, by virtue of (\ref{fourth}), this metric is necessarily conformally flat.

\medskip

{\bf Remark.} Notice that any Euler-Lagrange equation $(f_{u_x})_x+ (f_{u_y})_y+
(f_{u_t})_t=0$ is manifestly conservative and, moreover,  possesses three extra conservation laws
$$
\begin{array}{c}
(u_xf_{u_x}-f)_x+(u_xf_{u_y})_y+(u_xf_{u_t})_t=0, \\
\ \\
(u_yf_{u_x})_x+(u_yf_{u_y}-f)_y+(u_yf_{u_t})_t=0, \\
\ \\
(u_tf_{u_x})_x+(u_tf_{u_y})_y+(u_tf_{u_t}-f)_t=0,
\end{array}
$$
which are components of the energy-momentum tensor. In the dKP case,
$f=u_x^3/3+u_y^2-u_xu_t$,
these four conservation laws  take the form
$$
\begin{array}{c}
(u_x^2-u_t)_x+(2u_y)_y-(u_x)_t=0, \\
\ \\
(2u_x^3/3-u_y^2)_x+(2u_xu_y)_y-(u_x^2)_t=0, \\
\ \\
(u_x^2u_y-u_yu_t)_x+(u_xu_t+u_y^2-u_x^3/3)_y-(u_xu_y)_t=0, \\
\ \\
(u_x^2u_t-u_t^2)_x+(2u_yu_t)_y-(u_y^2+u_x^3/3)_t=0, \\
\end{array}
$$
respectively. One can show that the dKP equation possesses no extra conservation
laws of the form
$$
g(u_x, u_y, u_t)_x+ h(u_x, u_y, u_t)_y+p(u_x, u_y, u_t)_t=0.
$$
The discussion of the corresponding hierarchies of {\it higher nonlocal}
symmetries and conservation laws  for integrable Euler-Lagrange equations is beyond the scope of this paper. Our primary goal is the characterization of integrable Lagrangians based on the method of hydrodynamic reductions.

 \section{Derivation of the integrability conditions: proof of Theorem~1}

 Introducing the variables $a=u_x, \ b=u_y, \ c=u_t$ one rewrites the
Euler-Lagrange equation (\ref{zero})  in the first order form
\begin{equation}
a_y=b_x, ~~~ a_t=c_x, ~~~ b_t=c_y, ~~~ (f_a)_x+(f_b)_y+(f_c)_t=0.
\label{1}
\end{equation}
The idea of the method of hydrodynamic reductions
is to look for solutions  of the system (\ref{1}) in the form
$a=a(R^1, \ldots ,
R^n), \ b=b(R^1, \ldots , R^n), \ c=c(R^1, \ldots , R^n)$ where the  Riemann
invariants
$R^1(x,y,t)$, \ldots , $R^n(x,y,t)$ are {\it arbitrary} solutions of
 a pair of commuting hydrodynamic type flows
\begin{equation}
 R^i_y=\mu^i(R)\ R^i_x,  ~~~~ R^i_t=\lambda^i(R)\ R^i_x.
\label{R}
\end{equation}
Substituting this ansatz into (\ref{1}) one obtains the equations
\begin{equation}
\partial_ib=\mu^i\partial_ia, ~~~ \partial_ic=\lambda^i\partial_ia
\label{bc}
\end{equation}
(here $\partial_i=\partial/\partial_{R^i}$) along with the dispersion relation
\begin{equation}
D(\lambda^i, \mu^i)
=f_{aa}+2f_{ab}\mu^i+2f_{ac}\lambda^i+f_{bb}(\mu^i)^2+2f_{bc}\mu^i\lambda^i+f_{c
c}(\lambda^i)^2=0.
\label{disp}
\end{equation}
Hereafter, we assume the conic (\ref{disp}) to be irreducible. This condition is
equivalent to the
non-degeneracy of the Lagrangian or, equivalently, to the non-vanishing of the
Hessian: $H \ne 0$,  see~(\ref{Hessian}).
The consistency conditions of the equations (\ref{bc}) imply
\begin{equation}
\partial_i\partial_ja=\frac{\partial_j\lambda
^i}{\lambda^j-\lambda^i}\partial_i
a+\frac{\partial_i\lambda
^j}{\lambda^i-\lambda^j}\partial_j a.
\label{a}
\end{equation}
Differentiating the dispersion relation (\ref{disp}) with respect to $R^j, \
j\ne i,$ and keeping in mind
(\ref{bc}) and (\ref{comm}) one obtains the explicit expressions for $\partial_j
\lambda^i$ and
$\partial_j \mu^i$ in the form
\begin{equation}
\partial_j\lambda^i=(\lambda^i-\lambda^j)B_{ij}\partial_ja, ~~~
\partial_j\mu^i=(\mu^i-\mu^j)B_{ij}\partial_ja
\label{2}
\end{equation}
where $B_{ij}$ are  rational expressions in $\lambda^i, \lambda^j, \mu^i, \mu^j$
whose coefficients depend on partial derivatives of the density $f(a, b, c)$ up
to  third order. Explicitly, one has
$$
B_{ij}=\frac{N_{ij}}{D_{ij}}=\frac{N_{ij}}
{2(f_{aa}+f_{ab}(\mu^i+\mu^j)+f_{ac}(\lambda^i+\lambda^j)+f_{bb}\mu^i\mu^j+
f_{bc}(\mu^i\lambda^j+\mu^j\lambda^i)+f_{cc}\lambda^i\lambda^j)};
$$
notice that,  modulo the dispersion relation (\ref{disp}),  the denominator
$D_{ij}$ equals
$4D\left(\frac{\lambda^i+\lambda^j}{2}, \ \frac{\mu^i+\mu^j}{2}\right)$.
The numerator $N_{ij}$ is a polynomial expression of the form
$$
\begin{array}{c}
N_{ij}=f_{aaa}+f_{aab}(\mu^j+2\mu^i)+f_{aac}(\lambda^j+2\lambda^i)+
f_{abb}\mu^i(\mu^i+2\mu^j)+f_{acc}\lambda^i(\lambda^i+2\lambda^j)+ \\
\ \\
2f_{abc}(\lambda^i\mu^j+\lambda^j\mu^i+\lambda^i\mu^i)+f_{bbb}(\mu^i)^2\mu^j+f_{
ccc}(\lambda^i)^2\lambda^j+ \\
\ \\
f_{bbc}\mu^i(\lambda^j\mu^i+2\lambda^i\mu^j)+
f_{bcc}\lambda^i(\lambda^i\mu^j+2\lambda^j\mu^i).
\end{array}
$$
The equation (\ref{a}) takes the form
\begin{equation}
\partial_i\partial_ja=-(B_{ij}+B_{ji})\partial_ia\partial_ja.
\label{a1}
\end{equation}
The compatibility conditions $\partial_k\partial_j
\lambda^i=\partial_j\partial_k \lambda^i$,
 $\partial_k\partial_j \mu^i=\partial_j\partial_k \mu^i$ and
$\partial_k\partial_j\partial_i a=\partial_j\partial_k\partial_i a$ are
equivalent to the equations
\begin{equation}
\partial_kB_{ij}=(B_{ij}B_{kj}+B_{ij}B_{ik}-B_{kj}B_{ik})\partial_ka,
\label{int}
\end{equation}
which must be satisfied identically by virtue of (\ref{bc}), (\ref{disp}),
(\ref{2}).  The details of this calculation (which is computationally intense) can be summarized as follows.

In order to obtain equations with `smallest possible' coefficients at the fourth order
derivatives of $f(a,b,c)$ we rewrite (\ref{int}) as
\begin{equation}
\label{Nij}
\partial_kN_{ij}=N_{ij}\left(\frac{1}{D_{ij}}\partial_kD_{ij} +
B_{kj}\partial_ka+B_{ik}\partial_ka\right)-D_{ij}B_{kj}B_{ik}\partial_ka.
\end{equation}
The fourth order derivatives of $f(a,b,c)$ are present only in
the l.h.s.\ term $\partial_kN_{ij}$. Further reduction of the complexity
of the expression in the r.h.s.\ is achieved by representing
$1/D_{ij}$ in the form
$$
\begin{array}{c}
\displaystyle \frac{1}{D_{ij}} = U_{ij}=
[2(\lambda^if_{bc}+f_{ab})(\lambda^jf_{bc}+f_{ab})
  -f_{bb}(\lambda^i\lambda^jf_{cc}
 + (\lambda^i+\lambda^j)f_{ac} +f_{aa}) \\
  \ \\
+f_{bb}(\lambda^jf_{bc}+f_{ab})\mu^i
+f_{bb}(\lambda^if_{bc}+f_{ab})\mu^j
+f_{bb}^2\mu^i\mu^j ]
/( (\lambda^i- \lambda^j)^2 H)
\end{array}
$$
(which holds identically modulo the dispersion relation
(\ref{disp})), and the subsequent substitution
$B_{st}=N_{st}/D_{st} = N_{st}U_{st}$. The denominators of the
r.h.s.\ terms in (\ref{Nij}) cancel out as explained in the
attached program file 2-Lagr3dim.frm   producing a polynomial in
$\lambda^i$, $\lambda^j$, $\lambda^k$, $\mu^i$, $\mu^j$, $\mu^k$
with coefficients depending on the derivatives of the density
$f(a,b,c)$. This was the most essential technical part of the
calculation: the starting expression for the r.h.s.\ of
(\ref{Nij}) has more than 500.000 terms with different
denominators; after  properly organized cancellations we get a
polynomial expression with less than 2000 terms. Using
(\ref{disp}) and assuming $f_{bb} \neq 0$ (this can always be
achieved by a linear change of independent variables $x, y, t$),
we simplify this polynomial by substituting the powers of
$(\mu^i)^s$, $(\mu^j)^s$, $(\mu^k)^s$,  $s\geq 2$, arriving at a
polynomial of degree one in each of $\mu^i$, $\mu^j$, $\mu^k$ and
degree four in $\lambda$'s. Equating similar coefficients of these
polynomials in both sides of (\ref{Nij}) we arrive at a set of 45
equations for the derivatives of the Lagrangian density $f(a, b,
c)$ (linear in the fourth derivatives). Solving it, we get closed
form expressions for all fourth order derivatives of $f(a,b,c)$ in
terms of its second and third order derivatives. A straigtforward
computation shows that the compatibility conditions are satisfied
identically. Although the arising expressions are very long
indeed, they can be rewritten in a compact form (\ref{fourth}).
This finishes the proof of Theorem 1.

Further particulars and the programs in FORM \cite{FORM} and
Maple\footnote{Maple(TM) is a trademark of Waterloo Maple Inc.}
\cite{Maple} for the computations described above are given in the
attached files. The file 1-README explains the overall structure
of this program package.

{\bf Remark.} The integrability conditions (\ref{fourth}) are invariant
 under the obvious equivalence transformations generated by

\noindent (e1) linear transformations of $a, b, c$ corresponding to
linear changes of the independent variables $x, y, t$;

\noindent (e2) translations in $a, b, c$ corresponding to the
transformation $u\to u+\alpha x+ \beta y+\gamma t$;

\noindent (e3) transformations of the form
$f\to \mu f+\nu a +\eta b+\tau c
+\rho$ which do not effect the Euler-Lagrange equations.

These transformations generate a $17$-parameter symmetry group of the system
(\ref{fourth}). Notice that (e1) and (e2) generate the group of affine transformations. These symmetries allow one to considerably simplify the classification results. For instance, given any quadratic polynomial $Q(a, b, c)$ and a linear function $l(a, b, c)$, both not necessarily homogeneous, the density
$$
f(a, b, c)=Q(a, b, c)/l(a, b, c)
$$
is a solution of (\ref{fourth}). Up to the equivalence (e1) -- (e3) this expression can be transformed to either of the non-equivalent (over reals) canonical forms
$$
\frac{b^2+c^2+1}{a}, ~~~~ \frac{bc+1}{a}, ~~~~ \frac{b^2+c}{a}.
$$
The classification below is carried out up to this natural
equivalence.

\section{lntegrable polynomial Lagrangians}

We will start with a useful remark on homogeneous solutions of the equation (\ref{fourth}). Recall that a function $f$ is said to be homogeneous of degree $k$ if it satisfies the Euler identity
$af_a+bf_b+cf_c=kf$. This implies, in particular,  that $f_a, f_b, f_c$ are homogeneous of degree $k-1$, that is,
\begin{equation}
G\left(\begin{array}{c}
a \\
b\\
c
\end{array} \right)=(k-1)\left(\begin{array}{c}
f_a \\
f_b\\
f_c
\end{array} \right);
\label{id}
\end{equation}
here the $3\times 3$ matrix $G$ is the Hessian matrix of $f$.

\medskip

{\bf Lemma} {\it Let $f$ be a homogeneous solution of the equation (\ref{fourth}) with non-zero Hessian $H=det~G$. Then the degree of homogeneity $k$ can take only one of the three values $0, 2$ or $3$. }

\medskip

\centerline{Proof:}

The equation (\ref{fourth}) is an identity in $da, db, dc$. Let us replace $da, db, dc$ by $a, b, c$, respectively. Under this identification one has: $df\to kf, \ d^2f\to (k-1)k f, \ d^3f\to (k-2)(k-1)k f, \ d^4f\to (k-3)(k-2)(k-1)k f$. Moreover,
$$
\frac{dH}{H}=\frac{H_ada+H_bdb+H_cdc}{H}\to  \frac{H_aa+H_bb+H_cc}{H}=3(k-2),
$$
since $H$ is homogeneous of degree $3(k-2)$. Finally, the matrix one-form $dM$ reduces to
$$
\left(\begin{array}{cccc}
0 & (k-1)f_a & (k-1)f_b & (k-1)f_c \\
(k-1)f_a & (k-2)f_{aa} & (k-2)f_{ab} &(k-2) f_{ac} \\
(k-1)f_b &(k-2) f_{ab} & (k-2)f_{bb} &(k-2) f_{bc} \\
(k-1)f_c &(k-2) f_{ac} &(k-2) f_{bc} & (k-2)f_{cc}
\end{array}
\right);
$$
its determinant equals  $(k-1)^2(k-2)^2det~M$. Thus, for homogeneous $f$, the equation (\ref{fourth}) implies
\begin{equation}
(k-3)(k-2)(k-1)k f=3(k-2)^2(k-1)k f+\frac{3(k-1)^2(k-2)^2}{H}det~M.
\label{hom}
\end{equation}
We also point out the identity
$$
\frac{det~M}{H}=-
\left(\begin{array}{ccc}f_a& f_b &f_c\end{array}\right)\  G^{-1}\left(\begin{array}{c}
f_a \\
f_b\\
f_c
\end{array} \right).
$$
Taking into account (\ref{id}), this implies
$$
(k-1)\frac{det~M}{H}=-
\left(\begin{array}{ccc}f_a& f_b &f_c\end{array}\right)\left(\begin{array}{c}
a \\
b\\
c
\end{array} \right)=-kf.
$$
Thus, both terms on the right hand side of (\ref{hom}) cancel, leaving the identity $(k-3)(k-2)(k-1)k=0$.
It remains to point out that Lagrangians of homogeneity $k=1$ are automatically degenerate. Q.E.D.

\medskip

This result is particularly useful for the classification of polynomial solutions. Indeed, let us represent a polynomial  $f$ of degree $k$ in the form
$$
f=Q_k+Q_{k-1}+...
$$
where $Q_k$ is a {\it homogeneous} polynomial  of degree $k$ in the variables $a, b, c$. Writing the equation (\ref{fourth}) in the homogeneous form
$$
Hd^4f=d^3f dH+3 det (dM),
$$
we readily see that the leading term $Q_k$ must be a solution itself. This leads to two possibilities:
either $k=3$ (see Sect. 3.1) or, if $k\geq 4$,  the Hessian of $Q_k$ must vanish identically. In the last case the classical result \cite{Gordan} (see also \cite{Olver}, p. 234)  implies the existence of a linear change of variables $a, b, c$ after which $Q_k$ becomes a binary form, that is, a homogeneous function of two variables only (say, $a$ and $b$). This partial `separation of variables' considerably simplifies all calculations (see Sect. 3.2 for a complete analysis of the case $k=4$).

\subsection{Classification of cubic Lagrangians}

The integrability conditions (\ref{fourth}) provide a straightforward
classification of integrable cubic Lagrangians
with the densities
$$
f(a, b, c)=C(a, b, c)+\alpha a^2+\beta b^2+\gamma c^2+ \mu ab +\nu ac +\eta bc
$$
where $C$ is a homogeneous cubic form in $a, b, c$. Using equivalence
transformations
of the type (e1)  one can bring $C$ to a canonical form, thus simplifying the
analysis. Direct calculations using (\ref{fourth})
reveal that the cubic $C$ must necessarily be totally
reducible, that is, a product of three linear forms (this condition is
independent of the quadratic part). Thus, we have four cases to consider,
depending on the mutual position of the corresponding three lines.

\medskip

\noindent {\bf Case 1: Three lines in a general position.} Without any loss of
generality we can assume
$C=abc$. The substitution of  $f$ into the integrability conditions
(\ref{fourth}) readily implies $\alpha=\beta=\gamma=0$. Since the   remaining
constants $\mu, \nu, \eta$ can be eliminated by the equivalence transformations
(e2), we have just one integrable Lagrangian density in this class,
$$
f= u_xu_yu_t,
$$
with the corresponding Euler-Lagrange equation
\begin{equation}
u_tu_{xy}+u_yu_{xt}+u_xu_{yt}=0.
\label{3lines}
\end{equation}
Notice that this Lagrangian density is equivalent to
$$
f=u_x^3+u_y^3+u_t^3-3u_xu_yu_t;
$$
the corresponding (complex) linear change of variables $x, y, t$ can be
reconstructed from the factorization
$$
a^3+b^3+c^3-3abc=(a+b+c)(a+\epsilon b+\bar \epsilon c)(a+\bar \epsilon b+
\epsilon c), ~~~ \epsilon=-\frac{1}{2}+i\frac{\sqrt 3}{2};
$$
in this case two of the three lines are complex conjugate.
One can show that the equation (\ref{3lines}) possesses the Lax pair
\begin{equation}
S_t=q ({S_x}/{u_x})\ u_t, ~~~ S_y=r ({S_x}/{u_x})\ u_y,
\label{pr}
\end{equation}
where the functions  $q(s)$ and $r(s)$ ($s\equiv S_x/u_x$) satisfy a pair of
ODEs
$$
q'=\frac{q-r}{r-s}, ~~~ r'=\frac{r-q}{q-s}.
$$
This system can be integrated in elliptic functions as follows. Introducing
$f(s)=q(s)-s$ and $g(s)=r(s)-s$ one first rewrites the system in a
translation-invariant form
$$
f'=f/g-2, ~~~ g'=g/f-2
$$
or, equivalently,
$$
(fg)'=-(f+g), ~~~ (f+g)'=\frac{(f+g)^2}{fg}-6.
$$
Setting $fg=u$, $f+g=-u'$ one obtains a second order ODE for $u$, namely,
$uu''+(u')^2-6u=0$, whose integral is given in parametric form $u=\wp (p), \
s=\zeta (p)$ (that is, $u(s)$ is obtained by excluding  $p$ from these two
equations). Here $\wp (p)$ is the Weierstrass $\wp$-function,
$(\wp'(p))^2=4\wp^3(p)+\lambda^2$ (notice that $g_2=0, \ g_3=-\lambda^2$), and
$\zeta (p)$ is the corresponding zeta-function: $\zeta' (p)=-\wp (p)$. Thus, in
parametric form, $fg=\wp(p), \ f+g=\wp'(p)/\wp(p)$. The Lax pair
(\ref{pr}) can be rewritten in  parametric form as follows: adding and
multiplying  equations (\ref{pr}) we have
$$
\begin{array}{c}
\displaystyle
{\frac{S_t}{u_t}+\frac{S_y}{u_y}=q+r=2s+f+g=2\zeta(p)+\frac{\wp'(p)}{\wp(p)}},
\\
\ \\
\displaystyle {\frac{S_t}{u_t}\frac{S_y}{u_y}=qr=s^2+s(f+g)+fg=\zeta^2(p)+\zeta
(p)\frac{\wp'(p)}{\wp(p)}+\wp (p)}.
\end{array}
$$
Solving  these equations for $\frac{S_t}{u_t}$ and $\frac{S_y}{u_y}$ and keeping
in mind that
$\frac{S_x}{u_x}=s=\zeta(p)$, we arrive at  parametric equations
\begin{equation}
\frac{S_x}{u_x}=\zeta(p), ~~~
\frac{S_y}{u_y}=\zeta(p)+\frac{\wp'(p)+\lambda}{2\wp(p)}, ~~~
 \frac{S_t}{u_t}=\zeta(p)+\frac{\wp'(p)-\lambda}{2\wp(p)}.
 \label{z}
 \end{equation}
 Notice that these equations imply the algebraic identity
 $$
 \left(\frac{S_x}{u_x}-\frac{S_y}{u_y}\right)^2
\left(\frac{S_x}{u_x}-\frac{S_t}{u_t}\right)^2
  \left(\frac{S_y}{u_y}-\frac{S_t}{u_t}\right)^2=\lambda^2.
  $$

\medskip

\noindent {\bf Case 2: Three lines through a common point.} Without any loss of
generality we can assume $C=ab(a+b)$. The substitution of the corresponding $f$
into the integrability conditions (\ref{fourth}) implies $\gamma=\nu=\eta=0$ so
that there is no $c$-dependence. Thus, this case gives no non-trivial examples.

\medskip

\noindent {\bf Case 3: One double line.} Here $C=a^2b$, and the substitution
into the integrability conditions  implies $\gamma=\nu=0$. Translations in $b$
and $a$ eliminate $\alpha$ and $\mu$. Furthemore, the linear transformation $
\beta b+\eta c \to c$ reduces $f$ to the canonical form $f=bc+a^2b$. The
corresponding  Lagrangian density
$$
f=u_x^2u_y+u_yu_t
$$
generates the equation
\begin{equation}
u_{yt}+u_yu_{xx}+2u_xu_{xy}=0
\label{double}
\end{equation}
which,  up to  a rescaling,    is a particular form of the so-called
dispersionless $r$-Dym equation \cite{Bla}, \cite{Manas},
 $$
u_{yt}=\frac{3-r}{2-r}\left(\frac{1}{2-r}u_yu_{xx}-\frac{1}{1-r}u_xu_{xy}\right).
 $$
Indeed, for  $r=4/3$ this equation possesses the Lagrangian
$$
\int (u_yu_t-\frac{15}{4}u_x^2u_y) \, dx\,dy.
$$
The equation (\ref{double}) possesses the Lax pair
$$
S_t=-2u_xS_x+\frac{2}{5}S_x^{5/2}, ~~~ S_y=2u_yS_x^{-1/2}.
$$

\noindent {\bf Case 4: One triple line.} Here $C=a^3/3$, and the substitution
into the integrability conditions  implies a single constraint $4\beta
\gamma-\eta^2=0$. Eliminating $\alpha$ by a translation of $a$, one
arrives at the expression
$f=a^3/3+(\sqrt{\beta} b+\sqrt{\gamma }c)^2+ a(\mu b +\nu c).$ The linear change
of variables
$\sqrt{\beta} b+\sqrt{\gamma }c\to b, \  \mu b +\nu c \to -c$ reduces $f$ to the
canonical form
$f=a^3/3+b^2-ac$ which corresponds to the dKP density
$$
f=u_x^3/3+u_y^2-u_xu_t.
$$


Geometrically, the cases 3 and 4 can be viewed as degenerations of the case 1.
It would be interesting to perform these degenerations explicitly on the level
of the corresponding PDEs and  Lax pairs.

\subsection{Classification of fourth order Lagrangians}

According to the  Lemma, the leading  quartic part of the Lagrangian density $f$ must necessarily be degenerate and, thus, can be written as a form in two variables (say, $a$ and $b$) after an appropriate linear transformation. Since any homogeneous binary quartic  can be reduced to one of the five non-equivalent forms $a^4+\mu a^2b^2+b^4$, $a^2b(a+b)$, $a^2b^2$, $a^3b$ or $a^4$
(depending on the mutual location of its four roots), we have five cases to consider. A direct substitution into (\ref{fourth}) implies that the first four cases lead to inconsistency (whatever cubic and quadratic terms are), while the last case leads, up to the equivalence transformations (e1)-(e3), to a unique solution $f=a^4+2a^2c-ab-c^2$. The
corresponding  Lagrangian density
$$
f=u_x^4+2u_x^2u_t-u_xu_y-u_t^2
$$
generates the equation
$$
u_{xy}=-u_{tt}+2u_tu_{xx}+4u_xu_{xt}+6u_x^2u_{xx}.
$$
Up to a rescaling, this is  a particular case of the so-called $r$-th dispersionless modified KP equation
\cite{Bla}, \cite{Manas},
$$
u_{xy}=\frac{3-r}{(2-r)^2} u_{tt}-\frac{(3-r)(1-r)}{2-r}u_tu_{xx}-\frac{(3-r)r}{2-r}u_xu_{xt}-
\frac{(3-r)(1-r)}{2}u_x^2u_{xx},
$$
corresponding to the parameter value $r=2/3$ (for which the equation becomes manifestly Lagrangian).

A similar analysis reveals that the integrability conditions (\ref{fourth}) possess no polynomial solutions  of degree $5$. We conjecture that the four  examples listed above exhaust the list of polynomial integrable Lagrangians which are essentially three-dimensional.

\section{Pseudopotentials: proof of Theorem~2}

In this section we prove that any integrable non-degenerate
Euler-Lagrange equation (\ref{zero})
possesses a  dispersionless Lax pair. We look for a
pseudopotential $S$ governed by the equations
\begin{equation}
S_t=F(S_x, \, u_x, \, u_y, \, u_t), ~~~
S_y=G(S_x, \, u_x, \, u_y, \, u_t).
\label{Lax}
\end{equation}
Calculating the consistency condition $S_{ty}=S_{yt}$
and using the equation (\ref{zero}) we arrive at five
relations among $F$ and $G$,
\begin{equation}
\begin{array}{c}
F_{\xi}G_a-G_{\xi}F_a=\frac{f_{aa}}{f_{bb}}F_b, \\
\ \\
F_{\xi}G_b-G_{\xi}F_b=2\frac{f_{ab}}{f_{bb}}F_b-F_a, \\
\ \\
F_{\xi}G_c-G_{\xi}F_c=2\frac{f_{ac}}{f_{bb}}F_b+G_a, \\
\ \\
G_c+\frac{f_{cc}}{f_{bb}}F_b=0, ~~~ G_b+2\frac{f_{bc}}{f_{bb}}F_b-F_c=0.
\end{array}
\label{10}
\end{equation}
Here the auxiliary variable $\xi$ denotes $S_x$ and  $a=u_x$, $b=u_y$, $c=u_t$.
We also assume  $f_{bb}\ne 0$:
for a non-trivial Lagrangian, this can always be achieved by an admissible
linear change of $x$, $y$, $t$.

{\bf Remark.} One has to distinguish between `true' and `fake' pseudopotentials,
the latter satisfying equations of the form
\begin{equation}
S_t=p(a, S_x) c+q(a, S_x), ~~~ S_y=p(a, S_x) b+r(a, S_x),
\label{fake}
\end{equation}
where the functions $p(a, \xi),\  q(a, \xi), \ r(a, \xi)$ solve
the system
$$
p_a+pp_{\xi}=0, ~~  q_a+pq_{\xi}=0, ~~ r_a+pr_{\xi}=0.
$$
Equations (\ref{fake}) are automatically consistent by virtue of the relations
$a_y=b_x, \ a_t=c_x, \ b_t=c_y$  and, therefore, generate no non-trivial
PDEs. Such pseudopotentials can be ruled out, for instance, by the requirement
$F_b\ne 0$. We
 assume this hereafter. For a  `true' dispersionless Lax pair (\ref{Lax})
the compatibility conditions must be {\it equivalent} to the equation
(\ref{zero}). One can readily verify that the compatibility conditions for the
Lax pair (\ref{Lax}), (\ref{10})
with  $F_b\ne 0$ imply (\ref{zero}).

Equations (\ref{10}) imply the
expressions for the first derivatives of $G$ in terms of the first derivatives
of $F$,
\begin{equation}
\begin{array}{c}
G_c=-\frac{f_{cc}}{f_{bb}}F_b, \\
\ \\
 G_b=F_c-2\frac{f_{bc}}{f_{bb}}F_b, \\
 \ \\

G_a=-F_{\xi}F_b\frac{f_{cc}}{f_{bb}}-F_{\xi}F_c\left(\frac{F_c}{F_b}-2\frac{f_{b
c}}{f_{bb}}\right)
 +2\frac{f_{ab}}{f_{bb}}F_c-\frac{F_aF_c}{F_b}-2\frac{f_{ac}}{f_{bb}}F_b, \\
 \ \\
  G_{\xi}=F_{\xi}\left(\frac{F_c}{F_b}-2\frac{f_{bc}}{f_{bb}}\right)
 -2\frac{f_{ab}}{f_{bb}}+\frac{F_a}{F_b},
\end{array}
\label{FG}
\end{equation}
along with the following extra relation involving the first derivatives of $F$
only:
\begin{equation}
\begin{array}{c}
f_{cc}F_{\xi}^2F_b^2+F_{\xi}^2F_c(f_{bb}F_c-2f_{bc}F_b)+2f_{ac}F_{\xi}F_b^2+ \\
\ \\
2F_{\xi}F_a(f_{bb}F_c-f_{bc}F_b)+f_{bb}F_a^2+f_{aa}F_b^2-2f_{ab}F_b(F_{\xi}F_c+F
_a)=0.
\end{array}
\label{rel}
\end{equation}
To close this system one proceeds as follows. Differentiating the relation
(\ref{rel}) by $a, b, c, \xi$ and calculating the compatibility conditions of
the equations (\ref{FG}) (that is, $G_{cb}=G_{bc}, \ G_{ca}=G_{ac}, \ G_{c
\xi}=G_{\xi c}$, etc, six conditions altogether), one arrives at a linear
system of ten equations for the  second partial derivatives of $F$. Solving this
system (we skip the resulting expressions for the second order derivatives of $F$ due to their
complexity) and calculating the resulting compatibility conditions
$F_{aab}=F_{aba}$, etc,  one arrives at exactly  the same integrability
conditions (\ref{fourth}) as obtained  in Sect 2. from the requirement of the
existence of hydrodynamic reductions. This proves  Theorem 2.


The necessary details of this computation are explained in the
attached file 1-README. The main problem was to simplify the
expressions for  second order derivatives of $F$: the original
expressions found using Maple were extremely complicated and would
put the calculation of compatibility conditions
($F_{aab}=F_{aba}$, etc.) beyond our reach: for instance, the
number of terms in the expression for $F_{aa}$ was equal  to
18328, the corresponding number for $F_{ac}$ was 9045, etc. Thus,
the hypothetical expression for $F_{aab}$ would have approximately
$10^9$ terms. After taking  into consideration the identity
(\ref{rel}) we have simplified $F_{aa}$ to an expression with 367
terms only; respectively, $F_{ac}$  had 173 terms.  Verifying the
necessary compatibility conditions and their equivalence to
({\ref{fourth}) thus became feasible.

{\bf Remark.} The functions $F$ and $G$ which determine the Lax pair
(\ref{Lax})
depend on five arbitrary constants.
Indeed, the values of $G, F, F_a, F_b, F_c$ can be prescribed arbitrarily at
any initial point. On the other hand, equations for pseudopotentials remain
form-invariant under the transformations
$$
S \to \alpha S +\beta x + \gamma y +\delta t+ \mu u
$$
which allow one to eliminate this arbitrariness in the generic
situation.
Hence, for a given non-degenerate equation (\ref{zero}),
the dispersionless Lax pair is essentially unique.

\section{Differential-geometric aspects of the integrability conditions}

Since the integrability conditions (\ref{fourth}) are preserved by the transformations (e1)-(e3), they should be expressible in terms of the corresponding invariants. Among the simplest of these  invariants is the Hessian metric
$d^2f$.  We point out that {\it flat} Hessian metrics have been discussed recently in a series of publications \cite{Dubrovin, Mokhov, Totaro} in the context of equations of associativity of two-dimensional topological field theory.
Let us recall that, for a Hessian metric $d^2f$,  Christoffel's symbols and the curvature tensor take particularly simple forms,
$$
\Gamma^k_{ij}=\frac{1}{2}f^{kl}f_{lij}, ~~~ R_{ijkl}=\frac{1}{4}f^{mp}(f_{imk}f_{pjl}-f_{iml}f_{pjk});
$$
here $f(x^i, ..., x^n)$ is a function of $n$ variables, $f_{ij}, \ f_{ijk}$ denote partial derivatives of $f$, and the matrix $f^{ij}$ is the inverse of $f_{ij}$. Notice that the Riemann tensor does not contain fourth order derivatives of $f$. The Ricci tensor $R_{ij}$ and the scalar curvature $R$ are defined as
$R_{ij}=f^{mp}R_{pimj},  \ R=f^{ij}R_{ji}$. The main object of conformal geometry is the Schouten tensor $$
P_{ij}=\frac{1}{n-2}R_{ij}-\frac{1}{2(n-1)(n-2)}Rf_{ij}.
$$
In three dimensions $(n=3)$ the condition of conformal flatness is expressible in the form $\nabla_kP_{ij}=\nabla_jP_{ik}$ which is equivalent to the vanishing of the so-called Cotton tensor. Our main observation is the following

{\bf Proposition.} {\it For any density $f(a, b, c)$ satisfying  the integrability conditions (\ref{fourth}) the Hessian metric $d^2f$ is conformally flat. }

The proof is a straightforward computer calculation. For instance,
the density $f=abc$ generates the Hessian metric
$abc\left(\frac{da}{a}\frac{db}{b}+\frac{da}{a}\frac{dc}{c}+\frac{db}{b}\frac{dc}{c}\right)$
which is manifestly conformally flat.
 The attached program file Cotton.frm performs the general computation in FORM.

\medskip

We also present the equivalent tensorial formulation of the integrability conditions (\ref{fourth}):
\begin{equation}
f_{ijkl}={Sym} \left( f_{ijk}f^{pq}f_{pql}-\frac{3}{2}f_{pqi}f_{rsj}f_{mk}f_{nl}\epsilon^{prm}\epsilon^{qsn}\right).
\label{g}
\end{equation}
Here $\epsilon^{ijk}$ is the totally antisymmetric tensor  dual to the volume form of the Hessian metric $d^2f$ (so that $\epsilon^{123}=1/\sqrt H, \ \epsilon^{213}=-1/\sqrt H,$  etc),  and $Sym$ denotes the  total symmetrization:
$$
Sym \ T_{ijkl}=\frac{1}{4!}\sum_{\sigma \in S_4}T_{\sigma{(i)} \sigma{(j)} \sigma{(k)} \sigma{(l)}}.
$$
The formula (\ref{g}) can be rewritten in a completely invariant differential-geometric way as follows.
First of all, the Hessian metric $f_{ij}$ should be replaced by  a metric $g_{ij}$ and a flat  affine connection $\nabla$  such that
\begin{equation}
\nabla_kg_{ij}=\nabla_jg_{ik},
\label{nabla}
\end{equation}
indeed, the metric $g_{ij}$ assumes the Hessian form in the flat coordinates of $\nabla$. Let us define the 3-tensor $g_{ijk}=\nabla_kg_{ij}$ (which is automatically totally symmetric) and the antisymmetric tensor $\epsilon^{ijk}$ such that $\epsilon^{123}=1/\sqrt {det~g_{ij}}, \ \epsilon^{213}=-1/\sqrt {\det~g_{ij}},$  etc.
Then (\ref{g}) can be rewritten as
\begin{equation}
\nabla_k\nabla_l\ g_{ij}={Sym} \left( g_{ijk}g^{pq}g_{pql}-\frac{3}{2}g_{pqi}g_{rsj}g_{mk}g_{nl}\epsilon^{prm}\epsilon^{qsn}\right).
\label{g1}
\end{equation}
Thus, the differential-geometric object underlying the  classification of integrable Lagrangians is a pair consisting of a metric $g_{ij}$ and a flat affine connection $\nabla$ which satisfy the relations (\ref{nabla}), (\ref{g1}).

\section*{Acknowledgements}

We thank Dr.~Maxim Pavlov for drawing our attention
to the problem of integrability of multi-dimensional Lagrangians and numerous
helpful discussions. We also thank the Royal Society for their financial support
of SPT to Loughborough making this collaboration possible.

\end{document}